\documentstyle[12pt,epsf]{article} 
\newcommand{\be}{\begin{equation}}
\newcommand{\en}{\end{equation}}
\newcommand{\bea}{\begin{eqnarray}}
\newcommand{\ena}{\end{eqnarray}}

\newcommand{\hbo}{\hbox to 1 true cm {\hfill } }
\newcommand{\tr}{\hbox{tr}}

\begin{document}

\vglue 1truecm
  
\vbox{ UNITU-THEP-13/2000  DRAFT 
\hfill August 28, 2000
}
  
\vfil
\centerline{\large\bf Vortex dominance of the $0^+$ and $2^+$ glueball mass } 
\centerline{\large\bf in $SU(2)$ lattice gauge theory }
  
\bigskip
\centerline{ Kurt Langfeld, Alexandra Sch\"afke$^*$ } 
\vspace{1 true cm} 
\centerline{ Institut f\"ur Theoretische Physik, Universit\"at 
   T\"ubingen }
\centerline{D--72076 T\"ubingen, Germany}
  
\vfil
\begin{abstract}
The $c$-vortex ensembles are constructed by means of the recently 
proposed cooling method which gradually removes the $SU(2)/Z_2$ coset 
fields from the $SU(2)$ lattice configurations and which thus reveals 
the $Z_2$ vortex vacuum texture. Using Teper's blocking method, 
the screening masses of the $0^+$ and the $2^+$ glueball is calculated 
from these vortex ensembles and compared with the masses obtained 
from full configurations. The masses of either case agree within the 
achieved numerical accuracy of 10\%. As a byproduct, we find that 
the overlaps of the Teper operators with the glueball wavefunctions 
are significantly larger in the case of the $c$-vortex ensembles.

\end{abstract}

\vfil
\hrule width 5truecm
\vskip .2truecm
\begin{quote} 
$^*$ Supported in part by Graduiertenkolleg {\it Hadronen und Kerne.} 


keywords: {\it confinement, vortex dominance, glueball mass, SU(2) 
lattice gauge theory. } 

\end{quote}
\eject

{\bf Introduction.}
The idea that the center part of the $SU(N)$ gauge configurations 
are important for the confinement of quarks dates back to the 
late seventies~\cite{tho78}. Indeed, the proposal that the vortex free 
energy serves as an order parameter for quark (de-)confinement was 
recently confirmed in the case of a $SU(2)$ gauge group by a large scale 
lattice simulation~\cite{kov00}. Subsequently, Mach and collaborators 
explicitly constructed a vortex signature of the Yang-Mills vacuum from 
gauge invariant variables~\cite{mack}. It was observed at that time 
that random fluctuations of the vortex structure disorders the 
Wilson loop and, hence, provides quark confinement. With the increase 
of the computer performance in the mid eighties, many research 
efforts were devoted to substantiate this idea on a quantitative 
level~\cite{tom81,ale00}. One finds that projecting the full lattice 
configurations onto its vortex content reproduces the string tension 
of the static quark anti--quark potential. Recently, it was pointed 
out that not only the asymptotic behavior of the potential but also 
the short range part which is due to gluon exchange is reproduced if 
the plaquette is used for a definition of the vortex 
ensemble~\cite{fab99}. Moreover, the properties of the vortices 
arising from the plaquette projection technique strongly depend on the 
size of the lattice spacing thus rendering a continuum 
interpretation of the latter vortices cumbersome~\cite{fab99}. 

\vskip 0.3cm 
A significant upturn of the vortex picture of quark confinement occurred 
with the construction of the $p$-vortices which are defined after adopting 
the so-called center gauge~\cite{deb98} by projecting the gauge fixed 
link variables onto center elements~\cite{deb97,deb98}. 
The fact that Yang-Mills lattice configurations which were reduced to their 
$p$-vortex content reproduce the string tension is often referred to 
as center dominance of the string tension. Moreover, it was observed 
that the $p$-vortices are sensible degrees of freedom in the continuum 
limit~\cite{la98,deb98}: the (area) density of the $p$-vortices as well 
as their binary interactions extrapolate to the continuum. The 
$p$-vortex picture of the Yang-Mills ground state also provides an 
appealing explanation of the de-confinement phase transition at finite 
temperatures~\cite{la99}. 

\vskip 0,3cm 
It was pointed out that the center gauge fixing which is prior 
to identify the physical vortex structure might be plagued by 
a so-called practical Gribov problem~\cite{kov99}. In addition, it was 
observed that the vortex properties are quite sensitive to the finite 
size of the lattice volume~\cite{bor00}. For avoiding the practical 
Gribov and related problems, a gauge invariant definition of the vortex 
vacuum texture was achieved by employing a new self-restricted 
cooling procedure which diminishes the coset fields while leaving 
the center degrees of freedom un-changed~\cite{la00b}. For rating the 
phenomenological importance of these, say, $c$-vortices, center 
dominance of the string tension was verified. In addition, the $SU(2)$ 
action density which is carried by the $c$-vortex vacuum texture properly 
extrapolates to the continuum limit and, hence, gives rise to a mass 
dimension four condensate which features in the operator product 
expansion~\cite{la00b}. 

\vskip 0,3cm 
In this letter, we will investigate the question whether the $c$-vortex 
vacuum texture, besides the appealing picture for quark confinement and 
their hypothetical signature in high energy hadron collision 
experiments~\cite{la00b}, also provides a quantitative picture of the low 
lying excitations of pure Yang-Mills theory. For these purposes, we will 
calculate for the first time the correlation function for the 
$0^+$ and $2^+$ glueball channel, respectively, using vortex 
projected configurations. For this investigation, we confine 
ourselves to the most simple, but academic case of a pure $SU(2)$ gauge 
theory. An analogous investigation was recently performed using ablian 
projection to the maximum ablian gauge~\cite{sta99}: in this case, 
it was observed that abelian (or even monopole) projected configurations 
still reproduce the $0^+$/$2^+$glueball masses known from the full theory.

\vskip 0,3cm 
{\bf Glueball correlation functions.} 
Glueball screening masses $m_g$ are calculated from the 
correlation functions 
\be 
C(t) \; = \; \langle \widetilde{\phi }(t) \, \widetilde{\phi }(0) 
\rangle 
\; , \hbo 
\widetilde{\phi }(t) \; := \; \phi (t) \, - \, \langle \phi \rangle 
\label{eq:1} 
\en 
by analyzing the exponential decrease of $C(t)$ at asymptotic values 
of $t$, i.e. 
\be 
C(t) \; \propto \; \exp \biggl\{ - m_g t \biggr\} \; 
\hbox to 3cm {\hfill for \hfill } t \gg 1/m_g \; . 
\label{eq:2} 
\en 
Thereby $ \phi (x)$ is a combination of the link variables which 
carry the quantum numbers of the glueball under investigation. 
For choosing a function $ \phi (x)$ which generates sufficient overlap 
with the glueball state, we closely follow the pioneering work of 
Teper~\cite{tep86} and define composite link variables 
\bea 
U^{(N)}_i (x) &=& {\cal N} \biggl\{ 
U^{(N-1)}_i (x) U^{(N-1)}_i (x+i) 
\label{eq:3} \\ 
&+& \sum _{k \not= \pm i } 
U^{(N-1)}_k (x) U^{(N-1)}_i (x+k) U^{(N-1)}_i (x+k+i) 
U^{(N-1)\, \dagger }_k (x+i +i) \biggr\} \; , 
\nonumber 
\ena
where $i=1 \ldots 3, k \in {-3,-2,-1,1,2,3} $ and where 
${\cal N}$ is a normalization factor to ensure $U^{(N)}_i 
U^{(N) \, \dagger }_i  = 1$. 
The link variables $U^{(N)}_i (x)$ are defined on a coarser lattice 
of size $(N^{(N)}_s/2)^3 N_t$ where $N^{(N)}_s$ and $N_t$ are the number 
of lattice points in the spatial directions and in the time direction, 
respectively, of the finer lattice. The level $N=0$ corresponds 
to the finest level of a $N^{(0)}_s \equiv N_s$. 
The glueball operators are defined by means of the plaquette 
$P^{(N)}_{ik}(x)$ 
\bea 
\phi ^{0^+} (t) &=& \tr \sum _{\vec{x} } \biggl[ 
P^{(N)}_{12}(\vec{x},t) \, + \, P^{(N)}_{23}(\vec{x},t) \, + \, 
P^{(N)}_{13}(\vec{x},t) \biggr] \; , 
\label{eq:4} \\ 
\phi ^{2^+} (t) &=& \tr \sum _{\vec{x} } \biggl[ 
P^{(N)}_{12}(\vec{x},t) \, - \, P^{(N)}_{13}(\vec{x},t) \biggr] \; , 
\label{eq:5} 
\ena 
Here, we use the ''blocking'' level $N=2$ throughout the letter. 
Using the composite link method for the glueball operators, the glueball 
screening masses were successfully obtained even for the more realistic case 
of the gauge group $SU(3)$~\cite{mic89}. 

\vskip 0.3cm 
\begin{figure}[t]
\centerline{ 
\epsfxsize=7cm
\epsffile{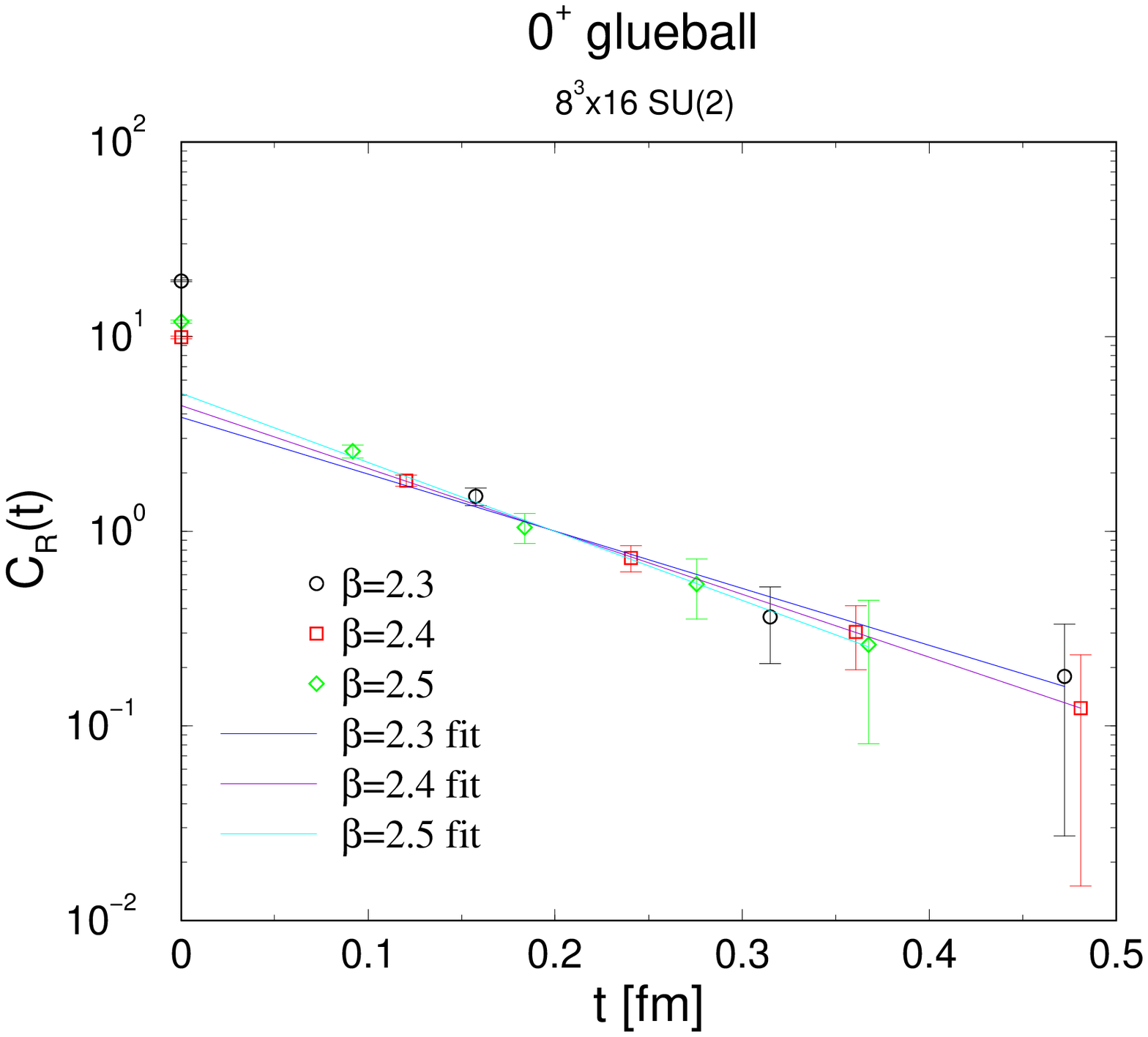}
\epsfxsize=7cm
\epsffile{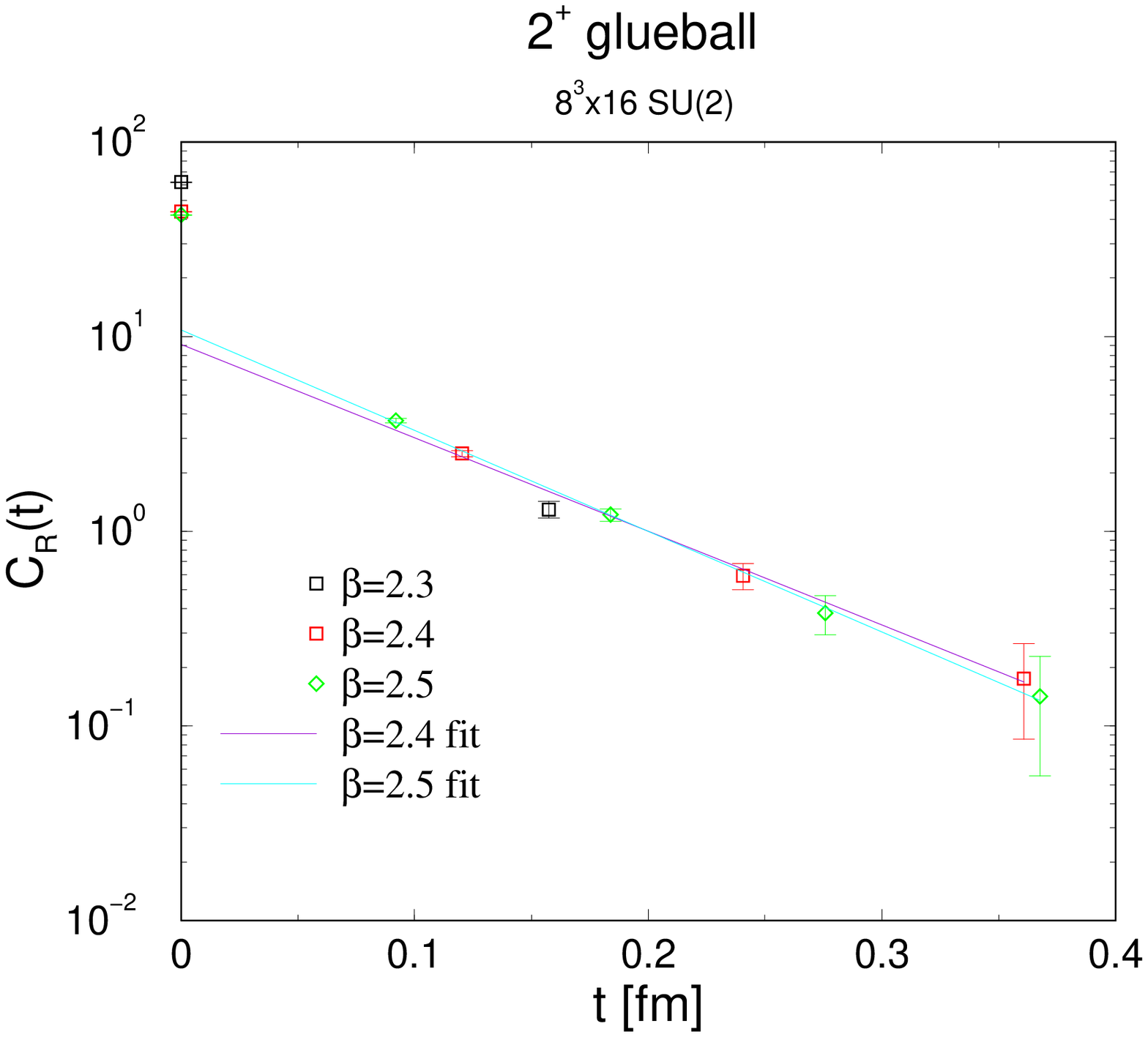}
}
\caption{ Renormalized correlator $C_R(t)$ (\ref{eq:6}) in the $0^+$ 
   and $2^+$ glueball channel, respectively, obtained from full 
   $SU(2)$ configurations. } 
\label{fig:1} 
\end{figure} 

In the present paper, we use a $N_s=8$, $N_t=16$ lattice and 
the standard heat bath algorithm to generate the 
link configurations according to a probability distribution provided by 
the Wilson action. Using $\beta = 4/g^2$ (where $g$ is the bare gauge 
coupling) up to values $2.5$, we are aware that finite size effects become 
visible. Our point is, however, to compare the glueball masses calculated 
from full configurations with those which were obtained by reducing 
the lattice variables to vortex ensembles rather than to perform 
a high precision extrapolation to the infinite volume limit. 
$15000$ measurements separated by $10$ Monte-Carlo sweeps to reduce
auto-correlations were performed to estimate $C(t)$ (\ref{eq:1}). 
In order to express $t$ in physical units, we use the $\beta $-dependence 
of the lattice spacing $a$ predicted by one-loop perturbation theory, i.e.
\be 
\sigma a^2(\beta ) \; = \; 0.12 \, \exp \biggl\{ - \frac{ 6 \pi ^2 }{11} 
(\beta - 2.3 ) \bigg\} \; , 
\label{eq:5a} 
\en 
where the string tension $\sigma = (440 \mathrm{MeV})^2 $ was used as 
reference scale. 

\vskip 0.3cm 
Since $C(0)$ is the expectation value of a composite field, $C(0)$ 
acquires additional divergencies even if $C(t\not=0)$ is 
renormalized finite~\cite{col77}. Hence, we refrain from normalizing 
$C(0)$ to $1$, but demand 
\be 
C_R(t) \; = \; Z^2_{\phi } \, C(t) \; , \hbo C_R(t_0) \; = \; 1 \; , 
\label{eq:6} 
\en 
where $t_0$ is the renormalization point. From our numerical simulation, 
the function $C(t)$ at the points $t=a,\ldots , \, 4a$ obeys the 
exponential law (\ref{eq:2}) to good accuracy. This fact allows 
to evaluate $C(t_0)$ by interpolation where $t_0 = 0.2 \,$fm is used 
throughout this paper. The final result $C_R(t)$ is shown in figure 
\ref{fig:1} for $\beta = 2.3, \, 2.4, \, 2.5$. The corresponding 
data points are satisfactorily close to a single exponential curve, 
thus establishing a renormalization group invariant screening mass. 
Note that $C_R(0)$ changes if different $\beta $ values are used, thus 
reflecting the additional divergency associated with the composite 
operator. The straight lines in figure (\ref{fig:1}) represent 
exponential fits to $C_R(t)$ for each $\beta $. Averaging 
over the screening masses obtained by the fit for a given $\beta $, 
we find 
\be 
m_{0^+} \; \approx \; 1.67 \pm 0.11 \, \mathrm{GeV} \; , \hbo 
m_{2^+} \; \approx \; 2.30 \pm 0.08 \, \mathrm{GeV} \; . 
\label{eq:7} 
\en 
The uncertainties provided in~(\ref{eq:7}) comprise statistical as 
well as systematic errors due to the extrapolation to the continuum 
limit. The masses (\ref{eq:7}) are consistent with the data presented 
in~\cite{tep86,sta99}. 

\vskip 0.3cm 
{\bf Glueball masses from the c-vortex ensembles.}  
In order to reveal the vortex vacuum structure, we employ 
the self-restricted cooling procedure proposed in~\cite{la00b} and 
fractionize the gauge group $SU(2) \hat{=} Z_2 \times SO(3)$. 
The corresponding degrees of freedom are center vortices and coset fields.
The coset part of the $SU(2)$ link variables $U_\mu (x)$ is isomorphic 
to the adjoint link 
\be
O^{ab}_\mu (x) \; = \; \frac{1}{2} \tr \biggl\{ U_\mu (x)~\tau ^a~
U^{\dagger}_\mu (x)~\tau^b \biggr\} \; = \; O^{ab}[A^b_{\mu}] \; ,
\hbo O^{ab}_\mu (x) \in SO(3) \; ,
\label{eq:9}
\en
which can be uniquely represented by a gauge vector potential $A_\mu (x)$
in the standard fashion. For removing 
gluonic (coset) degrees of freedom from $SU(2)$ configurations, 
the gluonic action density per link is defined by~\cite{la00b} 
\be
s^{gl}_\mu (x) \; = \; \sum_{\bar{\nu } \not=  \pm \mu } \left\{
1 \; - \; \frac{1}{3} \, \tr _A \,
O_{\mu \bar{\nu } }(x) \right\} \; = \; \frac{1}{3}
\sum _{\bar{\nu } \not= \pm \mu } F^a_{\mu \bar{\nu }} [A]
~F^a_{\mu \bar{\nu } }[A]~a^4 \, + \, {\cal O}(a^6) \; ,
\label{eq:10}
\en
where $O_{\mu\nu}(x)$ is the plaquette calculated in terms of
the $SO(3)$ link elements $O_\mu (x)$ (\ref{eq:9}).
The sum over $\bar{\nu}$ runs from $-4 \ldots 4$.
$F^a_{\mu\nu}[A]$
is the (continuum) field strength functional of the (continuum)
gluon fields $A_\mu(x)$. 

\begin{figure}[t]
\centerline{ 
\epsfxsize=7cm
\epsffile{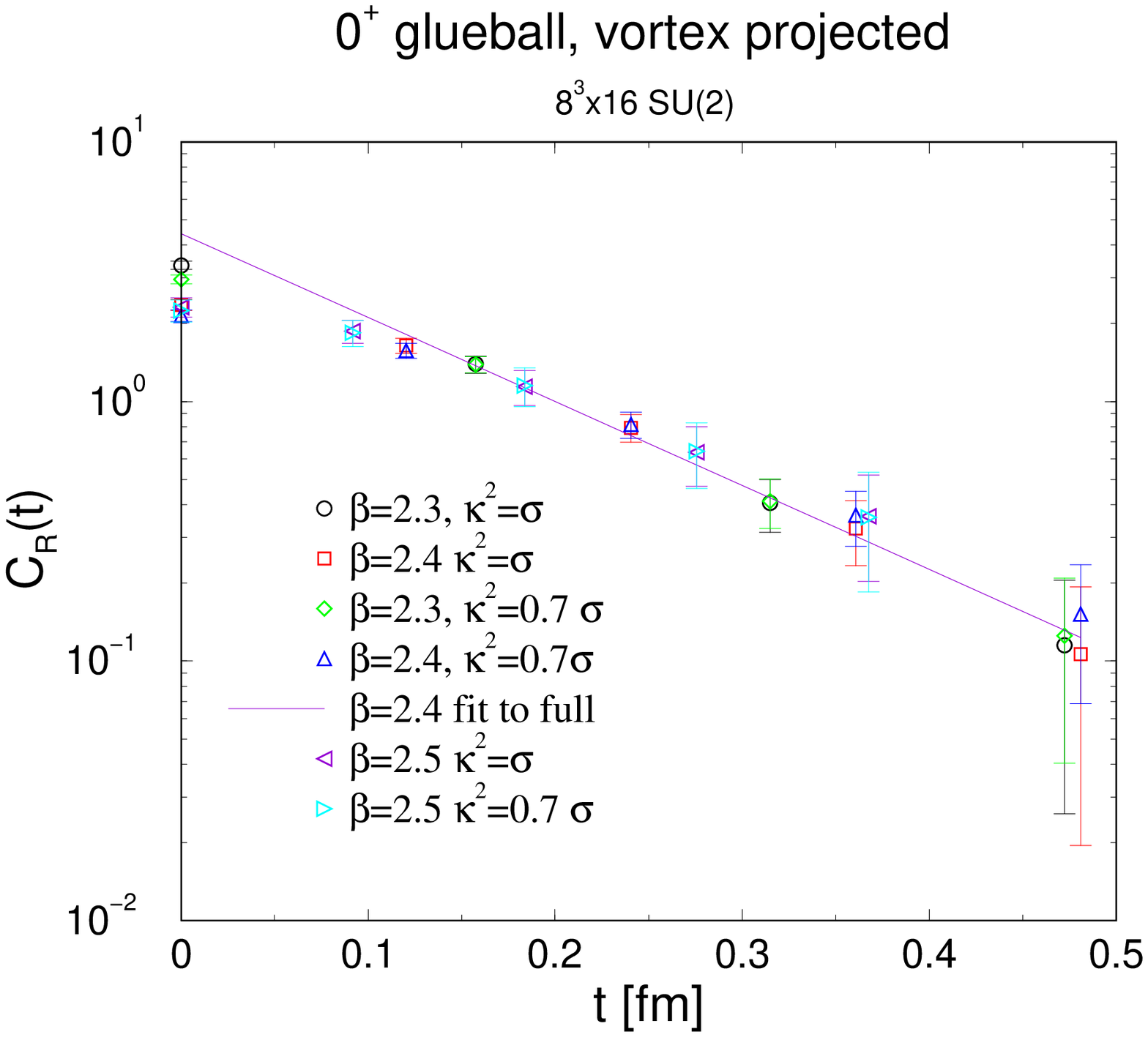}
\epsfxsize=7cm
\epsffile{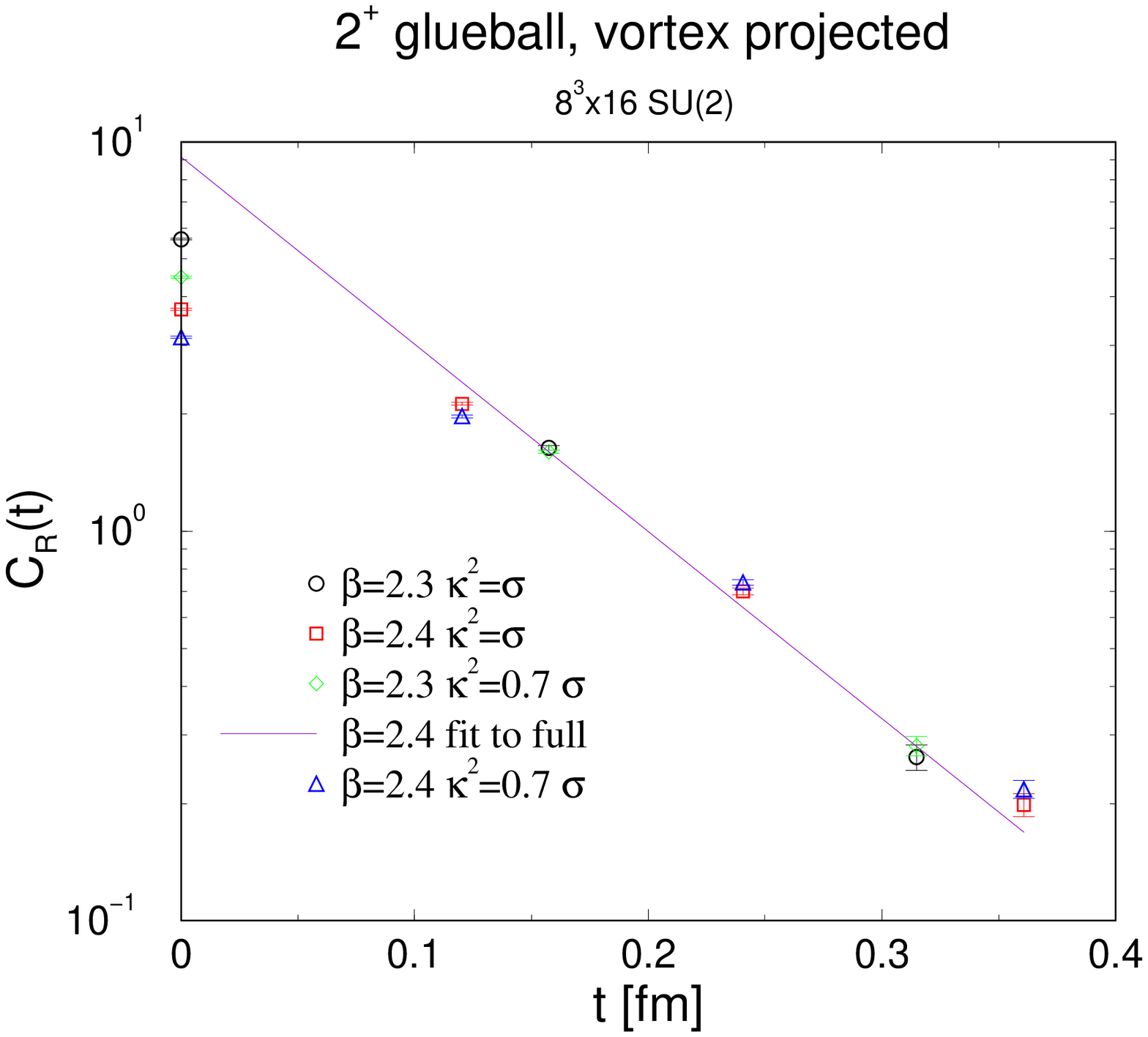}
}
\caption{ Renormalized correlator $C_R(t)$ (\ref{eq:6}) in the $0^+$ 
   and $2^+$ glueball channel, respectively, obtained  from $c$-vortex 
   ensembles.} 
\label{fig:2} 
\end{figure} 

\vskip 0.3cm
Cooling is performed by locally reducing the total gluonic action 
(\ref{eq:10}) with respect to the fields $O_\mu (x)$. 
Further cooling of the adjoint link $O_\mu (x)$ is rejected 
if the gluonic action is smaller than some threshold value
\be
s^{gl}_\mu (x) \; < \; 8 \kappa ^4 \, a^4 \; .
\label{eq:11}
\en
Thereby $\kappa $ is a gauge invariant cooling scale of mass
dimension one. The cooling procedure stops if the gluonic action 
density (\ref{eq:10}) locally has dropped below the critical value 
specified by $\kappa $. Details for the practical application of the 
cooling procedure can be found in~\cite{la00b}. 
For $\kappa=0$, the cooling procedure completely removes
the gluon fields from the SU(2) lattice configurations leaving only
gauge equivalents of $O_\mu (x)=1$. In fact, even for $\kappa ^2 
\approx \sigma $ the $SU(2)$ action density is largely clustered 
along 2-dimensional vortex world sheets, and the short range 
force between a static quark anti-quark pair is strongly affected. 
This is expected since the behavior at small distances is dominated by
the exchange of gluons, which are already partially eliminated by cooling.
Throughout this paper, the configurations which emerge after ''gluon'' 
cooling with $\kappa ^2 \le \sigma $ are labeled $c$-vortex ensembles. 

\vskip 0.3cm 
In order to investigate the $c$-vortex dominance of the glueball 
screening masses, we evaluate the glueball field combinations 
(\ref{eq:4}) and (\ref{eq:5}) using $c$-vortex ensembles obtained 
after adjoint cooling. Note that the action density of the 
gluon (coset) fields is by construction limited to $\kappa ^4 \le (440 \, 
\mathrm{MeV})^4$ (see (\ref{eq:10}) and (\ref{eq:11})) implying that the 
large $SU(2)$ action density carried by the $c$-vortices is required to 
sustain screening masses of order $1.5 \, $GeV. We employ $5000$ 
measurements to obtained the renormalized correlation function 
$C_R(t)$ calculated with adjoint cooled configurations, using 
$\kappa ^2 = \sigma $ and $\kappa ^2 = 0.5 \, \sigma $, respectively. 
The result is shown in figure~\ref{fig:2}. Again, the results for several 
values of $\beta $ are consistent with a single exponential law (\ref{eq:2}) 
reflecting proper scaling towards the continuum limit. 
The straight line shown in figure~\ref{fig:2} is the fit to the data 
obtained from {\it full} configurations for $\beta =2.4$ (see 
figure~\ref{fig:1}). We therefore find that the screening masses for 
the $0^+$/$2^+$ glueball calculated from full configurations and 
$c$-vortex ensembles, respectively, coincide within the 
achieved numerical accuracy. 

\vskip 0.3cm 
Note that the cooling procedure strongly effects the value 
$C_R(0)$. This is expected since the cooling procedure eliminates 
UV--divergencies which are generated by gluon (coset) fields 
of high ($SO(3)$) action density, and therefore alleviates the 
divergencies of the composite operator.  Furthermore, we point out that 
roughly the same statistical error of the screening masses 
was achieved in the case of the $c$-vortex ensembles with a number 
of measurements which is a factor of three less than the number 
of measurements employed in the case of full $SU(2)$ configurations. 
This improvement is due to an enhanced overlap of the ''wavefunctions'' 
(\ref{eq:4}) and (\ref{eq:5}) with the glueball wavefunctions 
once the adjoint cooling operates. Following Teper~\cite{tep86} 
for an estimate of the overlap, we find 
\be 
\frac{ C(2a) }{ C(0) } \biggl\vert _{\mathrm{vortex}} \; \approx \; 
4.6 \; \frac{ C(2a) }{ C(0) } \biggl\vert _{\mathrm{full}} \; , 
\hbo \beta =2.4, \, \kappa ^2 = \sigma 
\label{eq:12}
\en
for the $0^+$ glueball, and 
\be 
\frac{ C(2a) }{ C(0) } \biggl\vert _{\mathrm{vortex}} \; \approx \; 
13.9 \; \frac{ C(2a) }{ C(0) } \biggl\vert _{\mathrm{full}} \; , 
\hbo \beta =2.4, \, \kappa ^2 = \sigma 
\label{eq:13}
\en
for the $2^+$ glueball, respectively. 

\vskip 0.3cm 
{\bf Conclusions.} 
Recently, a self-restricted cooling method was proposed~\cite{la00b} 
which gradually removes the $SU(2)/Z_2$ gluon (coset) fields from
the $SU(2)$ lattice configurations paving the way to gauge invariant 
($c$-)vortex ensembles. Self-restriction ensures that the local 
gluonic action density does not exceed the cooling scale $\kappa $. 
By definition, $c$-vortex configurations are obtained for the choice 
$\kappa \le \sqrt{\sigma }$, where $\sigma = (440 \, \mathrm{MeV})^2$ 
is the string tension. In the present letter, we have studied the 
screening masses for the $0^+$ and $2^+$ glueballs extracted from 
Teper correlators~\cite{tep86}. Using $\kappa = \sqrt{\sigma } $ and 
$\kappa = 0.7 \, \sqrt{\sigma }$, we have shown that these screening masses 
are insensitive to the new cooling method hence providing evidence that 
the $0^+$ and $2^+$ glueball masses are dominated by $c$-vortex 
configurations.

\vspace{1cm}
{\bf Acknowledgements.} 
We thank H.~Reinhardt for encouragement and support.


\begin{thebibliography}{sch90}
\bibitem{tho78}{ G.~'t~Hooft, Nucl. Phys. {\bf B138} (1978) 1; \\ 
   Y.~Aharonov, A.~Casher and S.~Yankielowicz,
   Nucl. Phys. {\bf B146} (1978) 256; \\ 
   J.~M.~Cornwall, Nucl. Phys. {\bf B157} (1979) 392. } 
\bibitem{kov00}{ T.~G.~Kovacs and E.~T.~Tomboulis,
   Phys. Rev. Lett.  {\bf 85} (2000) 704. }
\bibitem{mack}{ G.~Mack and V.~B.~Petkova, Ann. Phys. (NY) {\bf 123}
   (1979) 442; \\
   G.~Mack, Phys. Rev. Lett. {\bf 45} (1980) 1378; \\
   G.~Mack and V.~B.~Petkova, Ann. Phys. (NY) {\bf 125} (1980) 117; \\
   G.~Mack, in: {\em Recent Developments in Gauge Theories},
   eds. G.~'t~Hooft et al. (Plenum, New York, 1980); \\
   G.~Mack and E.~Pietarinen, Nucl. Phys. {\bf B205} [FS5] (1982) 141. } 
\bibitem{tom81}{ E.~T.~Tomboulis, Phys. Rev. {\bf D 23} (1981) 2371; \\
   E.~T.~Tomboulis, Phys. Lett. {\bf B303} (1993) 103; \\ 
   T.~G.~Kovacs and E.~T.~Tomboulis,
   Nucl. Phys. Proc. Suppl.{\bf 63} (1998) 534. } 
\bibitem{ale00}{ A.~Alexandru and R.~W.~Haymaker,
   {\it Vortices in $SO(3)\times Z(2)$ simulations}, hep-lat/0002031. } 
\bibitem{fab99}{ M.~Faber, J.~Greensite and \v{S}.~Olejn{\'\i}k,
   JHEP {\bf 9901} (1999) 008; \\ 
   M.~C.~Ogilvie, Phys. Rev. {\bf D59} (1999) 074505. }
\bibitem{deb98}{ L.~Del Debbio, M.~Faber, J.~Giedt, J.~Greensite and
   \v{S}.~Olejn{\'\i}k, Phys. Rev. {\bf D 58} (1998) 094501. } 
\bibitem{deb97}{ L.~Del Debbio, M.~Faber, J.~Greensite and
   \v{S}.~Olejn{\'\i}k, Nucl. Phys. Proc. Suppl. {\bf 53} (1997) 141; \\ 
   L.~Del Debbio, M.~Faber, J.~Greensite and \v{S}.~Olejn{\'\i}k, 
   Phys. Rev. {\bf D 55} (1997) 2298. } 
\bibitem{la98}{ K.~Langfeld, H.~Reinhardt and O.~Tennert,
   Phys. Lett. {\bf B419} (1998) 317; 
   M.~Engelhardt, K.~Langfeld, H.~Reinhardt and O.~Tennert,
   Phys. Lett. {\bf B431} (1998) 141. } 
\bibitem{la99}{ K.~Langfeld, O.~Tennert, M.~Engelhardt and H.~Reinhardt,
   Phys. Lett. {\bf B452} (1999) 301; \\ 
   M.~Engelhardt, K.~Langfeld, H.~Reinhardt and O.~Tennert,
   Phys. Rev. {\bf D 61} (2000) 054504; \\ 
   J.~Gattnar, K.~Langfeld, A.~Sch\"afke and H.~Reinhardt,
   {\it Center-vortex dominance after dimensional reduction of SU(2) 
   lattice  gauge theory}, hep-lat/0005016, in press by Phys. Lett. B.} 
\bibitem{kov99}{ T.~G.~Kovacs and E.~T.~Tomboulis,
   Phys. Lett.  {\bf B463} (1999) 104. } 
\bibitem{bor00}{ V.~G.~Bornyakov, D.~A.~Komarov, M.~I.~Polikarpov 
   and A.~I.~Veselov, JETP Lett. {\bf 71} (2000) 231; \\ 
   R.~Bertle, M.~Faber, J.~Greensite and \v{S}.~Olejn{\'\i}k,
   {\it P-vortices, gauge copies, and lattice size}, hep-lat/0007043. } 
\bibitem{la00b}{K.~Langfeld, E.-M.~Ilgenfritz, H.~Reinhardt, 
   {\it Gauge invariant vortex vacuum textures and the gluon condensate}, 
   hep-lat/0008017; \\ 
   Kurt Langfeld, { The gluon condensate from gauge invariant vortex 
   vacuum texture}, contributed talk at "Confinement IV", July 3-8, 2000, 
   Vienna, Austria. } 
\bibitem{sta99}{ J.~D.~Stack and R.~Filipczyk,
   Nucl. Phys. {\bf B546} (1999) 333. } 
\bibitem{tep86}{ M.~Teper, Phys. Lett. {\bf B183} (1986) 345; 
   Phys. Lett. {\bf B185} (1987) 121. } 
\bibitem{mic89}{ C.~Michael and M.~Teper,
   Nucl. Phys.  {\bf B314} (1989) 347. } 
\bibitem{col77}{ J.~C.~Collins, A.~Duncan and S.~D.~Joglekar,
   Phys. Rev. {\bf D16} (1977) 438. }




\end{thebibliography}
\end{document}